\documentclass[a4paper,12pt]{article}

\usepackage{art}

\title{Nonlinear Stochastic Model of Return matching to the data of New York and Vilnius Stock Exchanges}
\author{V. Gontis, A. Kononovičius}
\date{Institute of Theoretical Physics and Astronomy of Vilnius University, \href{mailto:gontis@itpa.lt}{\underline{gontis@itpa.lt}}}

\begin{document}

\maketitle

\begin{abstract}We scale and analyze the empirical data of return from New York and Vilnius stock exchanges matching it to the same nonlinear double stochastic model of return in financial market.
\end{abstract}

\section{Introduction}
Volatility clustering, evaluated through slowly decaying auto-correlations, Hurst effect or $1/f^\beta$  noise for absolute returns, is a characteristic property of most financial assets return time series \cite{Willinger1999}. Statistical analysis alone is not able to provide a definite answer for the presence or absence of long range dependence phenomenon in stock returns or volatility, unless economic mechanisms are proposed to understand the origin of such phenomena \cite{Willinger1999, Cont2005}. Whether results of statistical analysis correspond to long range dependence is a difficult question subject to an ongoing statistical debate \cite{Cont2005, Mikosch2003}. The agent based economic models \cite{Kirman2002, Lux2000} as well as stochastic models \cite{Borland2004,Queiros2007,Gontis2008,Gontis2010} exhibiting long range dependence phenomenon in volatility or trading volume are of grate interest and remain an active topic of research. 

The properties of stochastic multiplicative point processes have been investigated analytically and numerically and the formula for the power spectrum has been derived \cite{Gontis2004}, later the model has been related with the general form of the multiplicative stochastic differential equation \cite{Kaulakys2006,Kaulakys2009}. The extensive empirical analysis of the financial market data, supporting the idea that the long-range volatility correlations arise from trading activity, provides valuable background for further development of the long-ranged memory stochastic models \cite{Plerou2001,Gabaix2003}. The power law behaviour of the auto-regressive conditional duration process \cite{Sato2004} based on the random multiplicative process and its special case the self-modulation process \cite{Takayasu2003}, exhibiting $1/f$ fluctuations, supported the idea of stochastic modelling with a power law probability density function (PDF) and long memory. A stochastic model of trading activity based on an stochastic differential equation (SDE) driven Poisson-like process has been already presented in \cite{Gontis2008}. In the paper \cite{Gontis2010} we proposed a double stochastic model, which generates time series of the return with two power law statistics, i.e., the PDF and the power spectral density of absolute return, reproducing the empirical data for the one-minute trading return in the NYSE.

In this contribution we analyze empirical data from Vilnius Stock Exchange (VSE)\footnote{Here we abbreviate the official name NASDAQ OMX Vilnius Stock Exchange for the convenience} in comparison with NYSE and stochastic model proposed in \cite{Gontis2010}. At the same time we demonstrate the scaling of statistical properties with longer time window of return. 

\section{The double stochastic model of return in financial market}
\label{sec:model}
Recently we proposed the double stochastic model of return in financial market \cite{Gontis2010} based on the nonlinear SDE. The main advantage of proposed model is its ability to reproduce power spectral density of absolute return as well as long term PDF of return. In the model proposed we assume that the empirical return $r_t$ can be written as instantaneous $q$-Gaussian fluctuations $\xi$ with a slowly diffusing parameter $r_0$ and constant $\lambda=5$%
\begin{equation} \label{eq:firstModel}
r_t = \xi \{ r_0, \lambda \} .
\end{equation}
q-Gaussian distribution of  can be written as follows:%
\begin{equation} 
P_{r_0,\lambda} (r) = \frac{\Gamma({\frac{\lambda}{2}})}{r_0 \sqrt{\pi} \Gamma({\frac{\lambda}{2}-\frac{1}{2}})} \left (\frac{r_0^2}{r_0^2+r^2} \right)^{\lambda/2} ,
\end{equation}
The parameter $r_0$ serves as a measure of instantaneous volatility of return fluctuations. See \cite{Gontis2010}, for the empirical evidence of this assumption. Here $r$ is defined in the selected time interval $\tau$ as a difference of logarithms of asset prices $p$:%
\begin{equation} 
r(t,\tau) = \left | \ln[p(t+\tau)] - \ln[p(t)] \right| .
\end{equation}
In this paper we consider dimensionless returns normalized by its dispersion calculated in the whole length of realization. It is worth to notice that $r(\tau)$ is an additive variable, i.e., if ${\tau=\sum\limits_i \tau_i}$, then $r(\tau)=\sum\limits_i r(\tau_i)$, or in the continuous limit the sum may be replaced by integration. We do propose to model the measure of volatility $r_0$  by the scaled continuous stochastic variable $x$, having a meaning of average return per unit time interval. By the empirical analyses of high frequency trading data on NYSE \cite{Gontis2010} we introduced relation:%
\begin{equation} 
r_0 (t,\tau)=1+\frac{\bar r_0}{\tau_s} \left | \int\limits_{t_s}^{t_s+\tau_s} x(s) \rmd s \right | ,
\end{equation}
where $\bar r_0$ is an empirical parameter and the average return per unit time interval $x(t_s)$ can be modeled by the nonlinear SDE, written in a scaled dimensionless time $t_s = \sigma_t^2 t$:%
\begin{equation} \label{eq:sde}
\rmd x = \left [ \eta - \frac{\lambda_0}{2} - \left (\frac{x}{x_{max}} \right)^2 \right ] \frac{\left (1+x^2 \right)^{\eta-1}}{(\epsilon \sqrt{1+x^2} +1)^2} x \rmd t_s + \frac{\left (1+x^2 \right)^{\frac{\eta}{2}}}{\epsilon \sqrt{1+x^2} +1} \rmd W_s.
\end{equation}
Here are five more empirically defined parameters: $\eta$ - exponent of multiplicativity, $\lambda_0$ - power law exponent of $x$ long range PDF, $\epsilon$ - parameter dividing diffusion into two areas: stationary and excited one, $\sigma_t^2$ - time scale adjustment parameter and $x_{max}$ - the upper limit of  diffusion. The term $ \left (\frac{x}{x_{max}} \right)^2$ excludes divergence of $x$ to the infinity. Seeking to discover the universal nature of financial markets we consider that all these parameters are universal for all stocks traded on various exchanges. In this paper we analyze empirical data from very different exchanges New York, one of the most developed with highly liquid stocks, and Vilnius, emerging one with stocks traded rarely.

We solve \eqref{sde} numerically introducing variable steps of dimensionless time ${t_{s,k+1}=t_{s,k}+h_k}$:%
\begin{equation}
h_k = \kappa^2 \frac{\left( \epsilon \sqrt{1+x_k^2} +1 \right)^2}{\left( 1+x_k^2\right)^{\eta-1}},
\end{equation}
where $\kappa$ is precision parameter of numerical calculations, which should be less than 1. Then SDE, \eqref{sde}, can be replaced by iterative equation:%
\begin{equation}\label{eq:lastModel}
x_{k+1} = \kappa^2 \left [ \eta - \frac{\lambda_0}{2} - \left (\frac{x}{x_{max}} \right)^2 \right ] x_k + \kappa \sqrt{1+x_k^2} \zeta_k ,
\end{equation}
where $\zeta_k$ is a normally distributed random variable with zero mean and unit variance.

\section{Stochastic model versus empirical data}

\begin{figure}
	\centering
	\includegraphics[width=\halfGraphWidth]{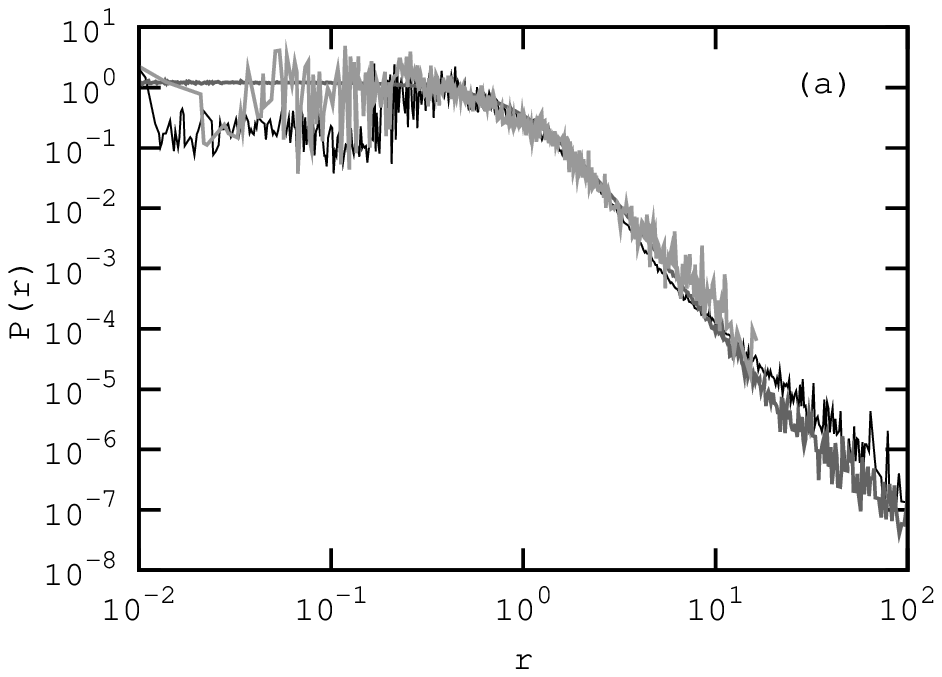}
	\hspace{\multiGraphSkip}
	\includegraphics[width=\halfGraphWidth]{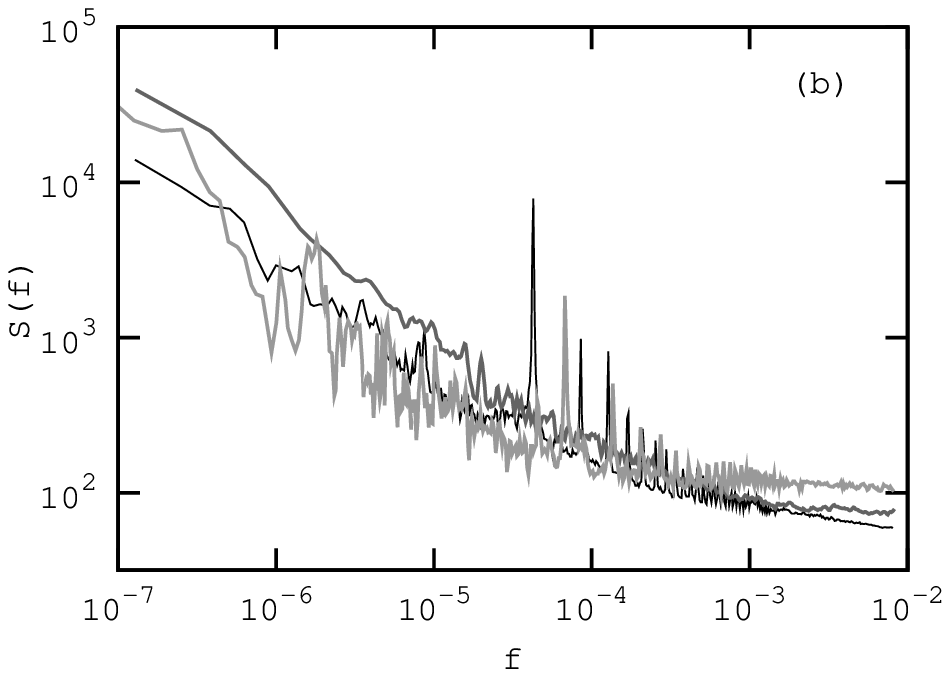}\\
	\includegraphics[width=\halfGraphWidth]{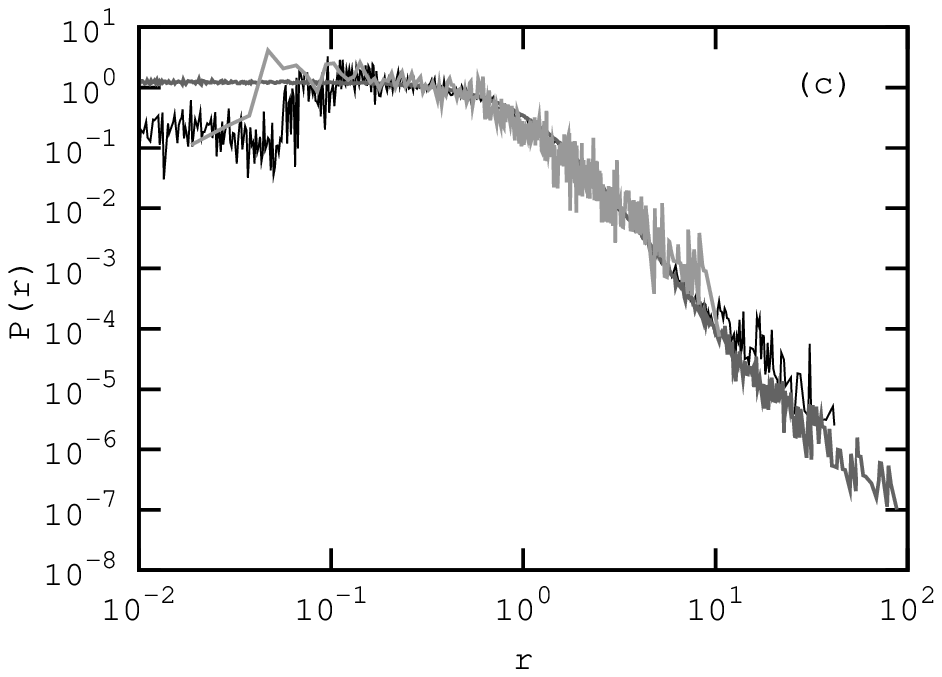}
	\hspace{\multiGraphSkip}
	\includegraphics[width=\halfGraphWidth]{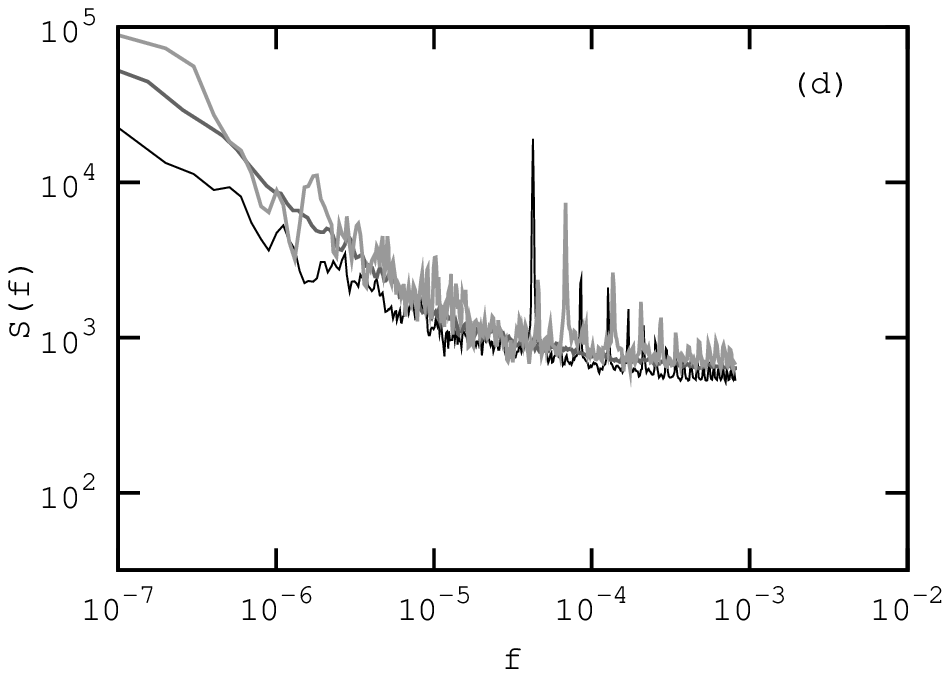}\\
	\includegraphics[width=\halfGraphWidth]{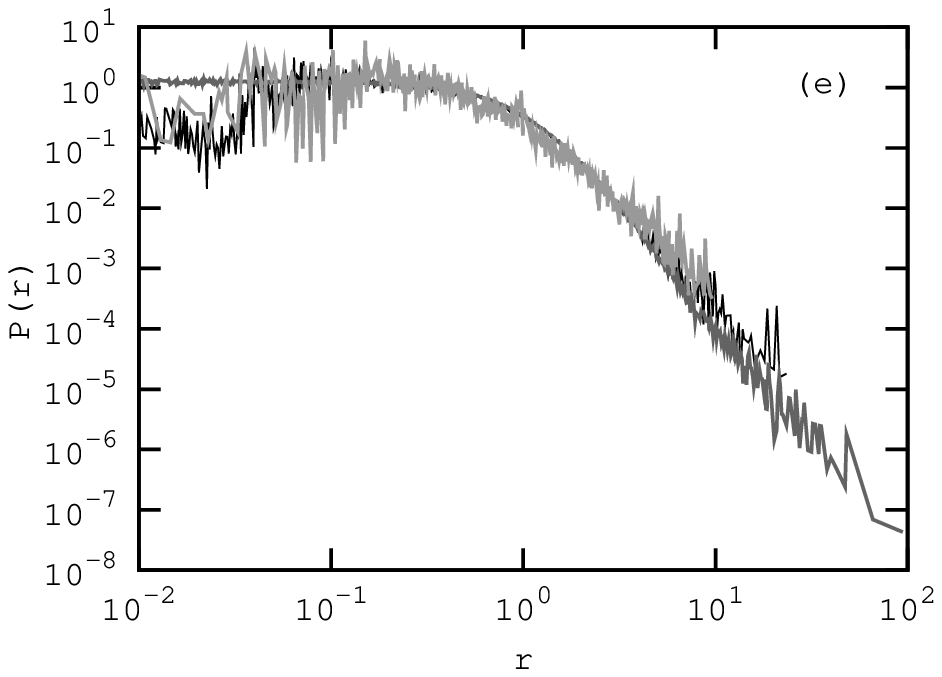}
	\hspace{\multiGraphSkip}
	\includegraphics[width=\halfGraphWidth]{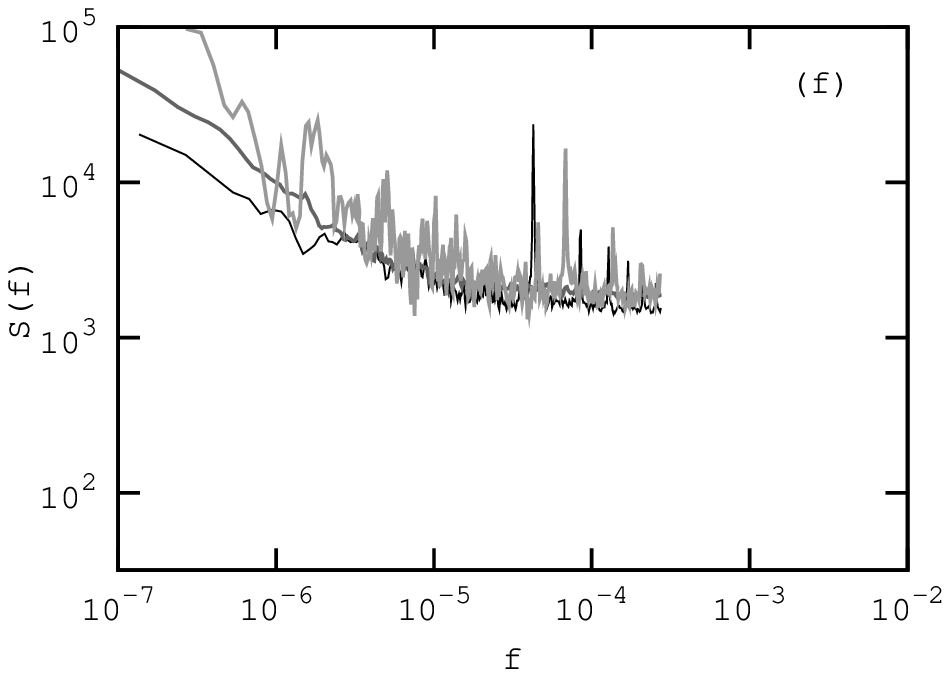}
	\caption{Comparison of empirical statistics of absolute returns traded on the NYSE (black thin lines) and VSE (light gray lines) with model statistics, \eqsref{firstModel}{lastModel}, (gray lines). Model parameters are as follows: $\lambda=5$; $\sigma_t^2=1/3 \cdot 10^{-6} s^{-1}$; $\lambda_0 = 3.6$; $\epsilon = 0.017$; $\eta = 2.5$; $\bar{r}_0 = 0.4$; $x_{max} = 1000$. PDF of normalized absolute returns is given on (a),(c),(e) and PSD on (b),(d),(f). (a) and (b) represents results with $\tau = 60 s$; (c) and (d) $\tau = 600 s$; (e) and (f) $\tau = 1800 s$. Empirical data from NYSE is averaged over 24 stocks and empirical data from VSE is averaged over 4 stocks.}
	\label{fig:comparison}
\end{figure}

In paper \cite{Gontis2010} we analyzed the tick by tick trades of 24 stocks, ABT, ADM, BMY, C, CVX, DOW, FNM, GE, GM, HD, IBM, JNJ, JPM, KO, LLY, MMM, MO, MOT, MRK, SLE, PFE, T, WMT, XOM, traded on the NYSE for 27 months from January, 2005, recorded in the Trades and Quotes database. The parameters of stochastic model presented in \secref{model} were adjusted to the empirical tick by tick one minute returns. An excellent agreement between empirical and model PDF and power spectrum was achieved, see Fig. 3 in \cite{Gontis2010}. The same empirical data and model results with slightly changed values of parameters are given in \figref{comparison} (a,b). Noticeable difference in theoretical and empirical PDFs for small values of return $r$ are related with the prevailing prices of trades expressed in integer values of cents. We do not account for this discreteness in our continuous description. In the empirical power spectrum one-day resonance - the largest spike with higher harmonics - is present. This seasonality - an intraday activity pattern of the signal - is not included in the model either and this leads to the explicable difference from observed power spectrum.

Provided that we use scaled dimensionless equations derived while making very general assumptions, we expect that proposed model should work for various assets traded on different exchanges as well as for various time scales $\tau$. We analyze tick by tick trades of 4 stocks, APG1L, PTR1L, SRS1L, UKB1L, traded on VSE for 50 months since May, 2005, trading data was collected and provided for us by VSE. Stocks traded on VSE in comparison with NYSE are less liquid -- mean inter-trade time for analyzed stocks traded on VSE is 362 s, while for stocks traded on NYSE mean inter-trade time equals 3.02 s. The difference in trading activity exceeds 100 times. This great difference is related with comparatively small number of traders and comparatively small companies participating in the emerging VSE market. Do these different markets have any statistical affinity is an essential question from the theoretical point of market modeling. 

First of all we start with returns for very small time scales $\tau = 60 s$. For the VSE up to 95\% of one minute trading time intervals elapse without any trade or price change. One can exclude these time intervals from the sequence calculating PDF of return. With such simple procedure calculated PDF of VSE empirical return overlaps with PDF of NYSE empirical return (see \figref{comparison} (a)).

One should use full time sequence of returns calculating the power spectrum. Nevertheless, despite low VSE liquidity, PSD of VSE and NYSE absolute returns almost overlap. Difference is clearly seen only for higher frequencies, when $\tau = 60 s$, and is related with low VSE market liquidity contributing to the white noise appearance. The different length of trading sessions in financial markets causes different positions of resonant spikes. One can conclude that even so marginal market as VSE retains essential statistical features as developed market on NYSE. At the first glance the statistical similarity should be even better for the higher values of return time scale $\tau$.

Further we investigate the behavior of returns on NYSE and VSE for increased values of $\tau = 600 s$ and $\tau = 1800 s$ with the specific interest to check whether proposed stochastic model scales in the same way as empirical data. Apparently, as we can see in \figref{comparison} (d) and (f) PSDs of absolute returns on VSE and on NYSE overlap even better at larger time scale ($600$ seconds and $1800$ seconds). This serves as an additional argument for the very general origin of long range memory properties observed in very different, liquidity-wise, markets. The nonlinear SDE is an applicable model to cache up observed empirical properties. PDFs of absolute return observed in both markets (see \figref{comparison} (c) and (e)) are practically identical, though we still have to ignore zero returns of VSE to arrive to the same normalization of PDF. 

\section{Conclusions}
We proposed a double stochastic process driven by the nonlinear scaled SDE \eqref{sde} reproducing the main statistical properties of the absolute return, observed in the financial markets. Seven parameters of the model enable us to adjust it to the sophisticated power law statistics of various stocks including long range behaviour. The scaled no dimensional form of equations gives an opportunity to deal with averaged statistics of various stocks and compare behaviour of different markets. All parameters introduced are recoverable from the empirical data and are responsible for the specific statistical features of real markets. Seeking to discover the universal nature of return statistics we analyse and compare extremely different markets in New York and Vilnius and adjust the model parameters to match statistics of both markets. The most promising result of this research is discovered increasing coincidence of the model with empirical data from the New York and Vilnius markets and between markets, when the time scale of return $\tau$ is growing. Observable specific features of different markets could be a subject of another research based on the proposed model. For example, it is clear that parameter $x_{max}$  should be relevant to the maximum number of active traders in the market and consequently should be specific for the every market. Further analyses of empirical data and proposed model reasoning by agent behavior is ongoing.

\sectionnonum{Acknowledgment}
We would like to express gratitude towards NASDAQ OMX Vilnius Stock Exchange, which provided empirical trade by trade data for our research.

\end{document}